\documentclass[
    ,final            
  ]
  {aipprocmod}

\layoutstyle{8x11single}

\usepackage{amsmath,amsthm,amssymb,amscd,mathrsfs}
\usepackage{wrapfig,url,slashed}

\makeatletter
\NAT@numberstrue
\def\NAT@cmprs{\@ne}

\newcommand{\special@letter}[1]{\mathrm{#1}}
\newcommand{\dd}{\special@letter{d}}
\newcommand{\ee}{\special@letter{e}}
\newcommand{\ii}{\special@letter{i}}
\newcommand{\DD}{\special@letter{D}}

\newcommand{\Order}{\special@letter{O}}

\DeclareMathAlphabet{\vcs}{OT1}{ptm}{b}{n}
\DeclareMathAlphabet{\vc}{OT1}{ptm}{b}{it}
\let\ol\overline

\newcommand{\s}[1]{_\mathrm{#1}}

\newcommand{\suprm}[1]{^\mathrm{#1}}

\newcommand{\EE}{\relax}
\def\EE{\@ifnextchar-{\@@EE}{\@EE}}
\def\@EE#1{\ifnum#1=1 \times\!10 \else \times\!10^{#1}\fi}
\def\@@EE#1#2{\times\!10^{-#2}}

\renewcommand{\tilde}{\expandafter\widetilde}

\newcommand{\un}[1]{\mbox{$\,\mathrm{#1}$}}

\newcommand{\Hu}{H\s u}  
\newcommand{\Hd}{H\s d}  
\newcommand{\bU}{{\bar U}} \newcommand{\bD}{{\bar D}} \newcommand{\bE}{{\bar E}}
 \newcommand{\bQ}{{\bar Q}}

\newcommand{\GeV}{\un{GeV}}
\newcommand{\TeV}{\un{TeV}}
\newcommand{\invfb}{\un{fb^{-1}}}
\def\thefigure{Fig.~\arabic{figure}}
\makeatother

\begin{document}

\title{Muon $g-2$ anomaly and 125\,GeV Higgs :\hfill\phantom{}\\
\hfill Extra vector-like quark and LHC prospects}


\author{Sho Iwamoto}{
  address={Department of Physics, University of Tokyo, Tokyo 113--0033, Japan}}

\begin{abstract}
The ATLAS and CMS collaborations recently reported indication of a Higgs
boson around 125\GeV. If we add extra vector-like quarks to the MSSM,
such a relatively heavy Higgs can be naturally realized in the GMSB framework,
simultaneously explaining the muon $g-2$ anomaly.
I will discuss LHC prospect of this attractive model.
\end{abstract}

\maketitle


\section{Introduction}
\paragraph{The Higgs boson}
The year 2011 was a great year for the Large Hadron Collider (LHC).
It delivered, and the ATLAS and CMS detectors respectively recorded, event data corresponds to c. $5\invfb$ with $\sqrt{s} = 7\TeV$.
The rich data yield tremendous development on the searches for the Standard Model (SM) Higgs boson; now the boson is likely to have a mass, if exists as we have hoped, within the range 116--130\GeV\ according to the ATLAS experiment, and 115--127\GeV\ to the CMS.

Our delight was that the both experiments have observed some excesses of events which can be interpreted as ``tantalizing hints'' of the Higgs boson with a mass of 124--126\GeV.
Such a Higgs boson mass is consistent with the prediction of the supersymmetry (SUSY), a promising candidate for a theory beyond the SM, which may lead us to the grand unified theory (GUT).

Under the minimal SUSY standard model (MSSM) framework, the Higgs boson mass\footnote{%
Here and hereafter, the ``Higgs boson'' refers to the lightest CP-even Higgs boson in the MSSM.}
 $m_h$ can be expressed with very rough approximation as
\begin{equation}
m_h^2\lesssim
m_Z^2 + \frac{3g_W^2m_t^4}{8\pi^2m_W^2}
\left[ \log\frac{m_{\tilde t}^2}{m_t^2} + \frac{X_t^2}{m_{\tilde t}^2}
\left(1-\frac1{12}\frac{X_t^2}{m_{\tilde t}^2}\right)\right],
\end{equation}
where $m^2_{\tilde t}:=(m_{\tilde t_1}^2+m_{\tilde t_2}^2)/2$ is the averaged mass of the top squarks, and $X_t:=A_t-\mu\cot\beta$ is a squark mixing parameter; $A_t$ is the SUSY-breaking trilinear coupling of the top squarks. 
Especially for the mass 125\GeV, we need $m_{\tilde t}$ more than a few TeV and/or rather large $X_t$, such as $X_t\approx -\sqrt6m_{\tilde t}$ (so-called $m_h$-max scenario)\footnote{For more rigid and quantitative discussions, see, e.g.~\cite{Ibe:2011aa,Draper:2011aa}.}.

\paragraph{Muon anomalous magnetic moment}
Now let us move on to another virtue of the SUSY: the explanation of the $(g-2)_\mu$ problem.
The muon anomalous magnetic moment $(g-2)_\mu$ has a discrepancy between its experimental result and the SM calculation; the values are, in terms of $a_\mu:=(g-2)_\mu/2$,
\begin{align}
 a_\mu\suprm{EXP} &= (11\,659\,208.9 \pm 6.3)\EE{-10}\quad\text{\cite{g-2_BNL2010}}&
 a_\mu\suprm{SM}  &= 
\begin{cases}
(11\,659\,208.9 \pm 6.3)\EE{-10}\quad\text{\cite{g-2_hagiwara2011}}\\
(11\,659\,180.2 \pm 4.9)\EE{-10}\quad\text{\cite{g-2_Davier2010}}
\end{cases}
\end{align}
where we can see that the discrepancy is a $3\sigma$ level or more.

\begin{figure}[t]\begin{center}
  \includegraphics[width=.48\textwidth]{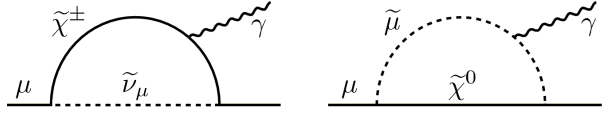}
  \caption{The SUSY diagrams contributing to the $(g-2)_\mu$ at the 1-loop level.}
  \label{fig:g-2}
\end{center}
\end{figure}

The SUSY has an ability to explain this discrepancy with its contribution to the $(g-2)_\mu$.
The contribution mainly comes from the two diagrams shown in \ref{fig:g-2}, and is approximately expressed as
\begin{align}
\Delta a_\mu\left(\tilde\chi^{\pm},\tilde\nu_\mu\right)&\approx \frac{\alpha\s w m_\mu^2}{m\s{soft}^2}\mathop{\mathrm{sgn}}(\mu M_2)\tan\beta,&
\Delta a_\mu\left(\tilde\chi^0,\tilde\mu\right)&\approx \frac{\alpha_Y m_\mu^2}{m\s{soft}^2}\mathop{\mathrm{sgn}}(\mu M_1)\tan\beta+\cdots.
\end{align}
Here $m\s{soft}$ is a representative mass of the relevant SUSY particles.
These expressions tell us that $m\s{soft}$ should be small, $\tan\beta$ be large, and $\mu$ be positive in order to realize the SUSY explanation of the $(g-2)_\mu$ problem.
Especially $m\s{soft}$ should be $\mathcal O(100\GeV)$.

\paragraph{A tug-of-war}
Now we can see a tension between the $125\GeV$ Higgs and the SUSY explanation of $(g-2)_\mu$; the former prefers a heavier SUSY, and the latter does a lighter.
Of course this is just a tension; we can realize the both simultaneously under the MSSM matter content with tuning the numerous SUSY-breaking parameters.
However, within simpler frameworks is the simultaneous realization impossible$^{\stepcounter{footnote}\thefootnote,\stepcounter{footnote}\thefootnote}$; for example, within the gauge-mediated SUSY breaking (GMSB)~\cite{Giudice:1998bp}, or minimal gravity-mediated SUSY breaking (mSUGRA) framework, which we have long loved.
\addtocounter{footnote}{-1}
\footnotetext{%
Checked by ourselves. For the mSUGRA case, see also Ref.~\cite{ghilencea:2012gz,endo:2011gy}.}
\stepcounter{footnote}
\footnotetext{%
This fact can be easily understood for the GMSB case.
In the framework $A_t$ tends to be very small, and thus we need $m_{\tilde t}\sim 10\TeV$.
Then the masses of the sleptons are also of order $1\TeV$, and $(g-2)_\mu$ cannot be explained.
}

Let us summarize our situation. Now we have two options to realize the $125\GeV$ Higgs with explaining the $(g-2)_\mu$ problem.
One choice is to abandon the GMSB or mSUGRA framework and consider more complex ones.
For example, once we set $m_0^{\rm 1st,\ 2nd}<m_0^{\rm 3rd}$ in the SUGRA framework,
i.e.~to set the scalar mass $m_0$ for the third family heavier than those for the first and second family, the simultaneous realization can be achieved\footnote{Checked by ourselves.}.
Another choice is to depart from the MSSM matter content with sticking to the GMSB/mSUGRA frameworks.
An example is the MSSM with an extra U(1) gauge symmetry~\cite{endo:2011gy}, and another example is the MSSM with vector-like matters~\cite{OkadaMoroi1992vectorlike,Babu2004vectorlike,Martin:2009bg,Endo2011vectorlike}, which we will investigate in the following part of this article.

Here it should be noted that the NMSSM does not work well. We need a large $\tan\beta$ for the $(g-2)_\mu$ explanation as we have seen, but increase of the Higgs boson mass due to the NMSSM system is sizable only if the $\tan\beta$ is very small ($\lesssim5$). Also $\text{MSSM}+\vcs{5}+\ol{\vcs{5}}$ model is inadequate; the Higgs boson mass is hardly increased~\cite{Martin:2009bg}.

\section{The Model}
In the model we consider~\cite{OkadaMoroi1992vectorlike,Babu2004vectorlike,Martin:2009bg,Endo2011vectorlike}, a vector-like pair of SU(5) complete multiplets $\vcs{10}+\ol{\vcs{10}}$ is attached to the MSSM as extra matters.
We denote the extra matters as $\vcs{10}=(Q',U',E')$ and $\ol{\vcs{10}}=(\bQ',\bU',\bE')$.
These  yield extra terms in the superpotential
\begin{align}
 W\s{add} &= Y' Q' \Hu U' + Y'' \bQ' \Hd \bU'  + m_{Q'} Q' \bQ' + m_{U'} U' \bU' + m_{E'} E' \bE',\\
 W\s{mix} &= \epsilon_i Q_i\Hu U' + \epsilon'_i Q'\Hu\bU_i + \epsilon''_i Q'\Hd\bD_i
             + \epsilon^L_i L_i\Hd\bE',
\end{align}
and corresponding soft terms.
Note that the mixing between the vector-like and the MSSM matters, now in $W\s{mix}$, is necessary, otherwise the extra matters result in stable charged particles.
This model is free from gauge anomalies, and does not spoil the gauge coupling unification~\cite{Babu2004vectorlike,Martin:2009bg}.

The Higgs boson mass in this model is lifted up by the extra top-like quark, especially due to the term $Y' Q' \Hu U'$. This term is quite similar to the term $y_t Q_3 \Hu U_3$ in the MSSM superpotential, and thus is capable to lift up the Higgs boson mass.
What is beautiful is that the coupling $Y'$ flows to an infrared fixed point $Y'\simeq 1.05$ through the renormalization group running. This means the lift-up due to the term becomes sizable without any assumptions.

However this model is still ugly in some viewpoints.
First of all, the coupling $Y''$ must be very small because it pulls down the Higgs mass with an opposite manner to $Y'$.
Another ugliness comes from the fact that the vector-like mass $M_{Q'}, M_{U'}$ must be around TeV scale as we will see later.
Furthermore, the mixing $\epsilon$'s must also be small, for a large mixing causes flavor--changing processes with unacceptable rates.
These features will be understood as assumptions in the following analysis.\footnote{%
Very recently, a model was proposed to solve the first two problems, i.e., smallness of $Y''$ and $m_{Q'}, m_{U'}$~\cite{Nakayama:2012zc}.
}

Finally, the increase of the Higgs boson mass can be approximately calculated with the 1-loop effective potential as, with assuming $Y''\ll Y'$ and $\tan\beta\gg1$,
\begin{equation}
\Delta m_h^2 \approx
\frac{3g_W^2 (Y'v)^4}{8\pi^2 m_w^2}\left[
\log\frac{M_S^2}{M_F^2}-\frac16\left(1-\frac{M_S^2}{M_F^2}\right)\left(5-\frac{M_S^2}{M_F^2}\right)
+\frac{A'^2}{M_S^2}\left(1-\frac{M_F^2}{3M_S^2}\right)-\frac1{12}\frac{A'^4}{M_S^4}
\right].
\end{equation}
Here $v=174\GeV$ is the vacuum expectation value of the Higgs boson, $M_F$ is a representative mass of the vector-like quarks, and $M_S$ is that of the vector-like squarks.
In other words, $M_S$ can be expressed with the soft mass $M\s{soft}$ as $M_S^2=M_F^2+M\s{soft}^2$.

It should be noted here that the extra vector-like quark does not increase the production rate of the Higgs boson~\cite{Ishiwata:2011hr}. Rather, the rate $gg\to h$ decreases by a few percent in this model.
This is because this model has two top-like extra quarks.
The heavier one does increase the rate $gg\to h$ by c. $\sim 10\%$, but the lighter one, whose contribution is larger than the heavier, decreases the rate $gg\to h$.

\begin{figure}[t]
\begin{center}
  \includegraphics[width=.75\textwidth]{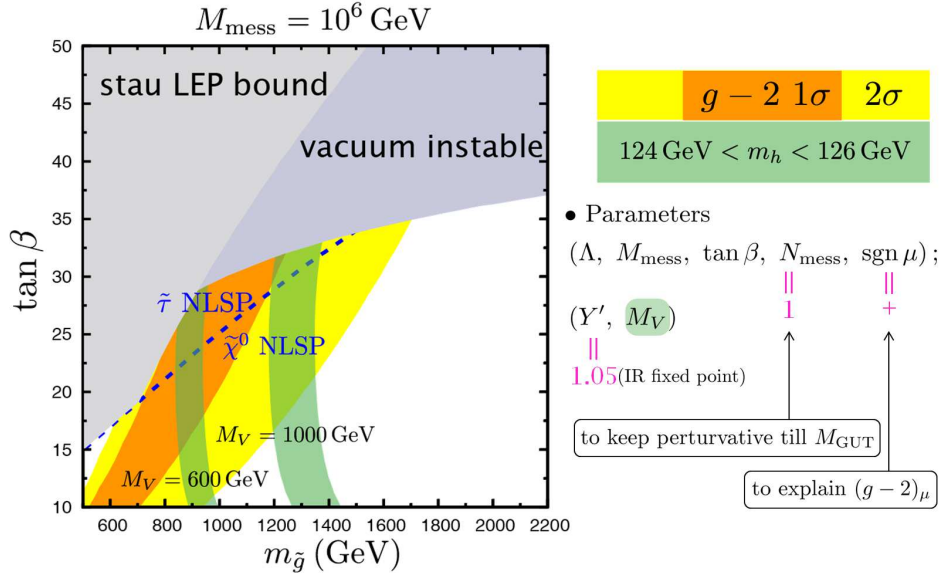}
  \caption{The conclusive panel of this article. The GMSB framework is considered, and the parameters are shown in the panel. The broad orange (yellow) band shows the region where the $(g-2)_\mu$ is within $1\sigma$ ($2\sigma$) from the experimental result.
  The two green bands are where the Higgs boson mass is within $[124,126]\GeV$; the left (right) one is for $M_V=600 (1000)\GeV$, where we set $m_{Q'}=m_{U'} (=:M_V)$.
  The LSP is the gravitino, and the NLSP is shown in the panel. Also two bounds are shown; one is the LEP bound of the stau search\cite{PDG2010}, and the other is from the condition that the lifetime of our vacuum should be larger than the age of the Universe\cite{Hisano:2010re,Endo:2012rd}.
}
  \label{fig:result}
\end{center}
\end{figure}
\section{Result and LHC prospects}
Here we focus on the GMSB case. The relevant parameters are the ordinal GMSB ones $(\Lambda, M\s{mess}, \tan\beta, N\s{mess}, \mathop{\mathrm{sgn}}\mu)$ and the extra ones $(m_{Q'}, m_{U'}, Y')$.
We set $m_{Q'}=m_{U'}=:M_V$ for simplicity, $\mu>0$ to explain the $(g-2)_\mu$ discrepancy, and $N\s{mess}=1$ to preserve the perturbativity of the gauge coupling constants up to the GUT scale.
The result is summarized in \ref{fig:result}, which is for $M\s{mess}=10^6\GeV$.
It can be read that the simultaneous realization is achieved around, e.g.,  $(m_{\tilde g},\tan\beta, M_V)=(1\TeV,20, 600\GeV)$.

However, with sadness, the great development of the LHC SUSY searches already excluded vast regions in the figure.
If the NLSP is a lighter stau $\tilde\tau_1$ and it is long-lived, the whole region shown in the figure (above the blue--dashed line) is rejected by the current bound $m_{\tilde \tau_1}>223\GeV$~\cite{Chatrchyan:2012sp}.
For other scenarios, the mass bound for the gluino $\tilde g$ excludes the left-hand-side region of the figure.
If the NLSP decays promptly, the bound is $m_{\tilde g}\gtrsim 1.2\TeV$ for a neutralino NLSP~\cite{Aad:2011zj} and $m_{\tilde g}\gtrsim 1.0\TeV$ for a stau NLSP~\cite{ATLAS:2012ag}.
The bound is somewhat relaxed for the case with a long-lived neutralino NLSP, but still it is $m_{\tilde g}\gtrsim 900\GeV$~\cite{ATLAS2012033,CMSPASSUS12005}.
A sizable region is still alive, but is barely breathing and the 2012 LHC run will draw a conclusion to the figure.

\begin{figure}[t]
\begin{center}
   \includegraphics[width=.45\textwidth]{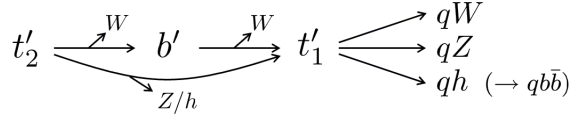}
  \caption{The decay tree of the vector-like quark.}
  \label{fig:decay}
\end{center}
\end{figure}

Searches for extra quarks are of great interest because the particles are specific and crucial to this model.
This model has three extra quarks (vector-like quarks), $t_1'$, $b'$, and $t_2'$ with masses
\begin{equation}
 m_{t_1'} \approx M_V - \frac v2, \qquad m_{b'} = M_V, \qquad m_{t_2'} \approx M_V+\frac v2,
\end{equation}
where $v=174\GeV$ is the vacuum expectation value of the Higgs and $M_V:=m_{Q'}=m_{U'}$ is assumed.
Because of the smallness of the mixing between the extra and SM quarks, the extra quarks are produced in pair at the LHC, and decays into the lightest one $t_1'$. Then $t_1'$ decays into $qW$, $qZ$ or $qh$ with a branching ratio determined by the mixing parameters $\epsilon_i$, $\epsilon_i'$, $\epsilon_i''$ in the superpotential $W\s{mix}$. (cf. \ref{fig:decay})

Mass bounds on the lightest vector-like quark $t_1'$ can be extracted from those on the heavy top-like quark $T$, but current bounds are obtained under the assumption that the particle $T$ has only one decay channel. For $T$ which decays exclusively as $T\to bW$, a bound $m_T>557\GeV$ is reported by the CMS collaboration~\cite{CMS:2012ab}. Similarly, a bound $m_T>350\GeV$ is obtained for $T$ with $T\to qW$~\cite{aaltonen:2011tq,aad:2012bt}, and $m_T > 475\GeV$ for $T\to tZ$~\cite{Chatrchyan:2011ay}.

The decay mode $t'_1\to qh$ is of great importance\footnote{%
Particles which decay into the Higgs boson are generally very curious targets, because they have a sizable coupling between the Higgs boson and thus would know about the Higgs sector to some extent.}. It can be utilized to the identification of the heavy top-like quark, and also it would carry information about the Higgs sector.
However, no result is reported on the search for the heavy top-like quark with respect to the decay channel $T\to qh$; it means in other words that we have no bound on $T$ which decays exclusively as $T\to qh$.
It shall be emphasized that such searches are now, on the edge of discovery of the Higgs boson, much anticipated.\footnote{Recently appeared a realistic study on possibility of the search for the heavy top-like quark focusing on the channel $T\to th$~\cite{harigaya:2012ir}. In the paper the top-like quark search with combining $T\to bW$ and $T\to th$ channels is investigated. It is also mentioned that the LHC data corresponds to $15\invfb$ with $\sqrt s = 8\TeV$ have capability to exclude the top-like quark less than 750 (650) GeV if the decay channel $T\to th$ $(T\to bW)$ is dominant.}

Searches for the heavier vector-like quarks, $b'$ and $t_2'$, are also interesting; with the searches we can distinguish $t'_1$ from the top-partner $t\s p$ which appears in the Little Higgs models.
Note that $t\s p$ has the same decay modes $t\s p\to th, tZ, bW$ as $t'_1$, while the chiral fourth generation quark only decays as $t_4 \to qW$.
However realistic studies on such searches should be come after the discovery of an extra top-like quark.

Searches for the vector-like lepton are also possible. Its decay branch might have information about the flavor structure of the leptons.
The signal will be clear, but for the smallness of the production cross section, the searches would require a lepton collider.

\section{Conclusion}
The MSSM--GMSB framework has attracted us because of its natural suppression of dangerous flavor--changing processes and CP violated ones.
However we cannot realize the $125\GeV$ Higgs boson and the $(g-2)_\mu$ explanation within the framework.
We have to depart from the simplest framework, or to add some extra matters to the MSSM.

We have seen that the extra vector-like matters $\vcs{10}+\ol{\vcs{10}}$ work very well.
They do increase the mass of the Higgs boson, and enable us to realize the $125\GeV$ Higgs boson with explaining the $(g-2)_\mu$ problem.

This model however suffers from the LHC SUSY search; we have faced that a vast region in \ref{fig:result} is under exclusion.
This is mainly because the explanation of the $(g-2)_\mu$ problem requires that the SUSY particles should be light at some level.
We would leastwise look forward to the result from the 2012 LHC run.

The searches for the extra vector-like particle, especially $t_1'$, has also been emphasized in this article.
It should be noted again that searches for the decay mode $T\to qh$ is much anticipated.

\begin{theacknowledgments}
  This article is based on the work~\cite{Endo2011vectorlike,Endo:2012rd} developed in collaboration with Dr.~Endo, Prof.~Hamaguchi and Dr.~Yokozaki. The author is very grateful to the collaborators.
  The author would like to thank the organizers of the GUT2012 Workshop at the Yukawa Institute for Theoretical Physics in Kyoto University for their warm hospitality and giving me the opportunity to have a talk.
  The author was supported by {\it JSPS Grant-in-Aid for JSPS Fellows}.
\end{theacknowledgments}


\bibliographystyle{mystylewotitle}   
\bibliography{proc_combined}


\end{document}